# Entanglement characteristics of subharmonic modes reflected from cavity for type II second harmonic generation

Zehui Zhai    Yongming Li    Jiangrui Gao[*]

State Key Laboratory of Quantum Optics and Quantum Optics Devices, Institute of Opto-Electronics, Shanxi University, Taiyuan, 030006, People's Republic of China

**Abstract**

Quantum fluctuation and quantum entanglement of the pump field reflected from an optical cavity for type II second harmonic generation are theoretically analyzed. The correlation spectra between the quadratures of the reflected subharmonic fields are interpreted in terms of pump parameter, intracavity losses and normalized frequency. Large correlation degrees of both amplitude and phase quadratures can be accessed in a triple resonant cavity before the pitchfork bifurcation occurs. The two reflected subharmonic fields are in an entangled state with the quantum correlation of phase quadratures and anticorrelation of amplitude quadratures. The proposed system can be exploited to be a new source generating entangled states of continuous variables.

## 1  Introduction

Quantum information has attracted great interest in recent years. The realization of quantum teleportation[1]-[4] and quantum dense coding[5]-[9] further enhanced confidence and passion in researching and developing quantum cryptography and quantum information processing. Quantum entanglement plays a central role in quantum information. The preparation of quantum entangled states is the most important basic work in the quantum information science. It is a significant research subject how to generate the entangled states using relatively simple systems.

[*] E mail: jrgao@sxu.edu.cn

The entanglement of quantum systems with continuous spectra is closely connected with squeezed states of optical fields. The bipartite entangled states of continuous electromagnetic fields have been generated from two single-mode squeezed vacuum states combined at a beam splitter and successfully applied in the experimental realizations of continuous-variable quantum teleportation of arbitrary coherent states[2]-[4]. The optical beams of Einstein-Podolsky-Rosen(EPR) entanglement have also been obtained by splitting a two-mode squeezed-state light with a polarizing beam splitter and used in the quantum optical communication[10] and dense coding[8]. Recently, by distributing one two-mode squeezed state among three partite using linear optics the bright tripartite entangled light beams are produced and utilized to achieve the controlled quantum dense coding[9]. A fully inseparable tripartite continuous-variable state is also obtained by combining three independent squeezed vacuum states[11]. So far, in all experiments the entangled states of continuous variables are generated from degenerate or nondegenerate optical parametric amplifiers (DOPA or NOPA) [2]-[13]. These experimental systems have to include two parts for the second harmonic generation (SHG) firstly and then the parametric down-conversion, respectively. Besides, it has been experimentally demonstrated that the pump fields reflected from an optical cavity for intracavity SHG are the squeezed-state light fields because of the cascaded nonlinear interaction between subharmonic and harmonic fields inside the cavity[14][15]. Z.Y.Ou[16] and C.Fabre et al.[15] theoretically analyzed the quantum fluctuation and squeezing characteristics of the reflected subharmonic fields and calculated the spectra of squeezing. It was pointed out in Ref. [16], for the case of type II harmonic generation there exists a threshold that is identified as the onset of an optical parametric oscillator (OPO) formed by a subharmonic mode with its polarization orthogonal to the input polarization (not the original modes) and the output of the orthogonal polarization mode from the OPO exhibits phase squeezing. M.W.Jack[17] et al. generalized the symmetric pumping case of Ref. [16] to asymmetric case which causes large changes in the classical dynamical behavior of the system. Squeezing and entanglement of doubly resonant type II SHG was analyzed in Ref.[18]. In Ref[16]-[18], the dependences of the quantum fluctuations and correlation variances of the quadratures of the reflected fields on the pump parameter($p/p_{th}$) involving with the pump power and the losses of the subharmonic modes are calculated, but the relation with the loss of the harmonic mode did not discussed.

In this paper, we analyze both classical behavior and the quantum correlations between the amplitude and phase quadratures of the two eigen subharmonic modes reflected from a single-ended cavity including a $\chi^{(2)}$ nonlinear crystal for type II second harmonic generation under the case of below the threshold. The pump field consisting of two frequency-degenerate subharmonic modes with orthogonal polarizations generates a harmonic field passing through an intracavity frequency-doubling process. The two subharmonic modes and the harmonic mode triply resonate in the optical cavity. The dependences of quantum correlations of the amplitude and phase quadratures on the analysis frequencies, the pump parameter and the loss of the harmonic field are numerically calculated. The results show that the two reflected subharmonic modes are in an entangled state with the phase-quadrature correlation and the amplitude-quadrature anticorrelation which has shown very important value in quantum communication[6][8][9]. It is found that the loss of harmonic wave strongly influences the quantum correlation in the case of the triple resonance with low loss. According to the Peres-Horodecki inseparability criterion of EPR entanglement state for continuous variables proposed by Duan[19], the inseparability between the two reflected subharmonic modes is confirmed by numerical calculations. The proposed system can be exploited to be a new type of entanglement sources for continuous variables with relatively simple configuration and high quantum correlation. The given numerical calculations may provide useful references for the design of the entanglement sources.

## 2  Fluctuation and correlation spectra of reflected pump modes

2.1 Equation of motion and its stationary solution

The sketch of SHG is shown in figure 1. We consider the SHG from a triply resonating optical cavity with a nonlinear $\chi^{(2)}$ crystal cut for type II phase matching. The triple resonance means that two nondegenerate pump modes with identical frequency and orthogonal polarizations and a harmonic mode simultaneously resonate in an optical cavity. Under the ideal case with perfect phase matching and without any detuning, the equations of motion for a single-ended cavity with one mirror used for input and output coupler can be expressed as [19]

$$\tau \dot{\alpha}_0(t) = -\gamma_0 \alpha_0(t) - \chi \alpha_1(t) \alpha_2(t) + \sqrt{2\gamma_0} c_0(t) \tag{1a}$$

$$\tau \dot{\alpha}_1(t) = -\gamma_1 \alpha_1(t) + \chi \alpha_2^*(t) \alpha_0(t) + \sqrt{2\gamma_{b1}} \alpha_1^{in} e^{i\phi_1} + \sqrt{2\gamma_1} c_1(t) \tag{1b}$$

$$\tau \dot{\alpha}_2(t) = -\gamma_2 \alpha_2(t) + \chi \alpha_1^*(t) \alpha_0(t) + \sqrt{2\gamma_{b2}} \alpha_2^{in} e^{i\phi_2} + \sqrt{2\gamma_{c2}} c_2(t). \tag{1c}$$

Where $\alpha_0, \alpha_1$ and $\alpha_2$ are the amplitudes of harmonic field and two pump fields inside the cavity respectively. The round-trip time of all modes in the cavity is assumed to be same. The single pass loss parameters $\gamma_{bi}$ and $\gamma_{ci}$ ($i=0,1,2$) stand for, respectively, the transmission through the input and output coupler and all extra loss mechanisms. $\gamma_i = \gamma_{bi} + \gamma_{ci}$ denotes the total loss coefficient and $\alpha_1^{in}, \alpha_2^{in}$ express the amplitudes of two input pump fields outside the coupler. For $\gamma_{bi} \ll 1$, the $\gamma_{bi}$ is related to the amplitude reflection coefficients $r_i$ and transmission coefficients $t_i$ of the coupler by the following formula:

$$r_i = 1 - \gamma_{bi}$$

$$t_i = \sqrt{2\gamma_{bi}}.$$

Assuming the two pump modes have the same positive real amplitude $\beta$, zero initial phase and the balanced loss in the cavity, we have

$$\gamma_1 = \gamma_2 = \gamma \tag{2a}$$

$$\gamma_{b1} = \gamma_{b2} = \gamma_b \tag{2b}$$

$$\gamma_{c1} = \gamma_{c2} = \gamma_c. \tag{2c}$$

The steady state solutions of equations (1) are obtained:

$$\bar{\alpha}_0 = \frac{-k\bar{\alpha}_1\bar{\alpha}_2}{\gamma_0} \tag{3a}$$

$$(-\gamma - \frac{\chi^2}{\gamma_0}|\bar{\alpha}_2|^2)\bar{\alpha}_1 + \sqrt{2\gamma_b}\beta = 0 \tag{3b}$$

$$(-\gamma - \frac{\chi^2}{\gamma_0}|\bar{\alpha}_1|^2)\bar{\alpha}_2 + \sqrt{2\gamma_b}\beta = 0, \tag{3c}$$

here $\bar{\alpha}_0$, $\bar{\alpha}_1$, $\bar{\alpha}_2$ are the steady-state amplitudes of the three intracavity modes $\alpha_0, \alpha_1$ and $\alpha_2$. The equations (3b) and (3c) show that both $\bar{\alpha}_1$ and $\bar{\alpha}_2$ are real numbers. The oscillation threshold $\beta^{th}$ and pump parameter $\sigma$ are expressed by

$$\beta^{th} = \sqrt{\frac{2\gamma^3\gamma_0}{\chi^2\gamma_b}} \tag{4a}$$

$$\sigma = \frac{\beta}{\beta^{th}}. \tag{4b}$$

Solving (3b) and (3c), the steady-state solutions of three modes above the threshold ($\sigma \geq 1$) are given by

$$\bar{\alpha}_1 = \frac{\sqrt{\gamma\gamma_0}}{\chi}\sigma - \frac{\sqrt{\gamma\gamma_0(\sigma^2-1)}}{\chi} \tag{5a}$$

$$\bar{\alpha}_2 = \frac{\sqrt{\gamma\gamma_0}}{\chi}\sigma + \frac{\sqrt{\gamma\gamma_0(\sigma^2-1)}}{\chi} \tag{5b}$$

$$\bar{\alpha}_0 = -\gamma/\chi, \tag{5c}$$

below the threshold ($\sigma \leq 1$), by

$$\bar{\alpha}_1 = \bar{\alpha}_2 = \alpha \tag{6a}$$

$$\alpha = \frac{\sqrt{\gamma\gamma_0}}{\chi}\sigma' \quad (6b)$$

$$\bar{\alpha}_0 = -\gamma\sigma'^2/\chi \quad (6c)$$

$$\sigma' = \left(\sigma + \sqrt{\sigma^2 + \frac{1}{27}}\right)^{1/3} - \frac{1}{3}\left(\sigma + \sqrt{\sigma^2 + \frac{1}{27}}\right)^{-1/3}.$$

**2.2, Correlation spectra of reflected subharmonic fields**

Entanglement characteristics of the two subharmonic modes reflected from the coupler are denoted with the correlations between the quantum fluctuations of their quadrature amplitude (X) and phase (Y) components. We consider the case of below threshold. The dynamics of the quantum fluctuations can be described by linearizing the classical equations of motion around the stationary state. Setting $\alpha_i = \bar{\alpha}_i + \delta\alpha_i$ (i=0,1,2), and using equation (1), we have

$$\tau\delta\dot{\alpha}_0(t) = -\gamma_0\delta\alpha_0(t) - \left[\chi\bar{\alpha}_2\delta\alpha_1(t) + \chi\bar{\alpha}_1\delta\alpha_2(t)\right] + \sqrt{2\gamma_0}c_0(t) \quad (7a)$$

$$\tau\delta\dot{\alpha}_1(t) = -\gamma\delta\alpha_1(t) + \chi\left[\bar{\alpha}_0\delta\alpha_2^*(t) + \bar{\alpha}_2\delta\alpha_0(t)\right] + \sqrt{2\gamma_b}b_1(t) + \sqrt{2\gamma_c}c_1(t) \quad (7b)$$

$$\tau\delta\dot{\alpha}_2(t) = -\gamma\delta\alpha_2(t) + \chi\left[\bar{\alpha}_0\delta\alpha_1^*(t) + \bar{\alpha}_1\delta\alpha_0(t)\right] + \sqrt{2\gamma_b}b_2(t) + \sqrt{2\gamma_c}c_2(t). \quad (7c)$$

The optical modes (O) can be expressed with amplitude quadratures (X) and phase quadratures (Y):

$$O = \frac{1}{2}(X + iY),$$

with $O = [\alpha_0, \alpha_1, \alpha_2, b_0, c_0, b_1, c_1, b_2, c_2]$, $X = [X_0, X_1, X_2, X_{b0}, X_{c0}, X_{b1}, X_{c1}, X_{b2}, X_{c2}]$ and $Y = [Y_0, Y_1, Y_2, Y_{b0}, Y_{c0}, Y_{b1}, Y_{c1}, Y_{b2}, Y_{c2}]$.

Substituting them into equations (7a)-(7c), we have

$$\tau\delta\dot{X}_0(t) = -\gamma_0\delta X_0(t) - \left[\chi\bar{\alpha}_2\delta X_1(t) + \chi\bar{\alpha}_1\delta X_2(t)\right] + \sqrt{2\gamma_0}\delta X_{c0}(t) \quad (8a)$$

$$\tau \delta \dot{X}_1(t) = -\gamma \delta X_1(t) + \chi \left[ \bar{\alpha}_0 \delta X_2(t) + \bar{\alpha} \delta X_0(t) \right] + \sqrt{2\gamma_b} \delta X_{b1}(t) + \sqrt{2\gamma_c} \delta X_{c1}(t) \quad (8b)$$

$$\tau \delta \dot{X}_2(t) = -\gamma \delta X_2(t) + \chi \left[ \bar{\alpha}_0 \delta X_1(t) + \bar{\alpha} \delta X_0(t) \right] + \sqrt{2\gamma_b} \delta X_{b2}(t) + \sqrt{2\gamma_c} \delta X_{c2}(t), \quad (8c)$$

and

$$\tau \delta \dot{Y}_0(t) = -\gamma_0 \delta Y_0(t) - \left[ \chi \bar{\alpha}_2 \delta Y_1(t) + \chi \bar{\alpha}_1 \delta Y_2(t) \right] + \sqrt{2\gamma_0} \delta Y_{c0}(t) \quad (9a)$$

$$\tau \delta \dot{Y}_1(t) = -\gamma \delta Y_1(t) - \chi \left[ \bar{\alpha}_0 \delta Y_2(t) - \bar{\alpha} \delta Y_0(t) \right] + \sqrt{2\gamma_b} \delta Y_{b1}(t) + \sqrt{2\gamma_c} \delta Y_{c1}(t) \quad (9b)$$

$$\tau \delta \dot{Y}_2(t) = -\gamma \delta Y_2(t) - \chi \left[ \bar{\alpha}_0 \delta Y_1(t) - \bar{\alpha} \delta Y_0(t) \right] + \sqrt{2\gamma_b} \delta Y_{b2}(t) + \sqrt{2\gamma_c} \delta Y_{c2}(t). \quad (9c)$$

Under the condition of below threshold, combining steady-state solution expressions (6a) and (6b), the correlation spectra $\delta X_1(\omega) + \delta X_2(\omega)$ and $\delta Y_1(\omega) - \delta Y_2(\omega)$ (Fourier transform at $\omega$), deduced from Eqs. (8) and (9) are given by

$$\delta X_1(\omega) + \delta X_2(\omega) = \frac{2\sqrt{\gamma\gamma_0}\sigma' Q_{x0} + (Q_{x1} + Q_{x2})(\gamma_0 + i\omega\tau)}{D} \quad (10a)$$

$$\delta Y_1(\omega) - \delta Y_2(\omega) = \frac{Q_{y1} - Q_{y2}}{\gamma + i\omega\tau + \gamma\sigma'^2}, \quad (10b)$$

where

$$D = (\gamma_0 + i\omega\tau)(\gamma + i\omega\tau + \gamma\sigma'^2) + 2\gamma\gamma_0\sigma'^2 \quad (11)$$

$$Q_{x(y)0} = \sqrt{2\gamma_0} \delta X(Y)_{c0} \quad (12a)$$

$$Q_{x(y)1} = \sqrt{2\gamma_b} \delta X(Y)_{b1} + \sqrt{2\gamma_c} \delta X(Y)_{c1} \quad (12b)$$

$$Q_{x(y)2} = \sqrt{2\gamma_b} \delta X(Y)_{b2} + \sqrt{2\gamma_c} \delta X(Y)_{c2}. \quad (12c)$$

Using the input-output relations of cavity:

$$\begin{aligned} \delta X_1^{out}(\omega) + \delta X_2^{out}(\omega) &= \sqrt{2\gamma_b}[\delta X_1(\omega) + \delta X_2(\omega)] - [\delta X_{b1}(\omega) + \delta X_{b21}(\omega)] \\ \delta Y_1^{out}(\omega) - \delta Y_2^{out}(\omega) &= \sqrt{2\gamma_b}[\delta Y_1(\omega) - \delta Y_2(\omega)] - [\delta Y_{b1}(\omega) - \delta Y_{b21}(\omega)] \end{aligned}, \quad (13)$$

the correlation spectra of the sum of amplitude quadratures and the difference of phase quadratures of outgoing subharmonic fields are obtained:

$$S_{X_1+X_2}^{out}(\Omega) = \left|\delta X_1^{out}(\omega) + \delta X_2^{out}(\omega)\right|^2$$
$$= \frac{8}{|D|^2}\left[2\left|\gamma_0\sigma'\sqrt{\gamma_b\gamma}\right|^2 + \left|\gamma_b(\gamma_0+i\Omega\gamma) - D/2\right|^2 + \left|\sqrt{\gamma_b\gamma_c}(\gamma_0+i\Omega\gamma)\right|^2\right] \quad (14a)$$

$$S_{Y_1-Y_2}^{out}(\Omega) = \left|\delta Y_1^{out}(\omega) - \delta Y_2^{out}(\omega)\right|^2$$
$$= 2\left|\frac{2\gamma_b}{\gamma+i\Omega\gamma+\gamma\sigma'2}-1\right|^2 + 2\left|\frac{2\sqrt{\gamma_b\gamma_c}}{\gamma+i\Omega\gamma+\gamma\sigma'2}\right|^2 \quad (14b)$$

where $\Omega = \frac{\omega\tau}{\gamma}$ is the normalized frequency.

Fig.2 shows the normalized correlation spectra of $S_{X_1+X_2}^{out}$ and $S_{Y_1-Y_2}^{out}$, both of which are smaller than the standard quantum limit (SQL, normalized to 1). Fig.3 shows the correlation variances of amplitude and phase quadrature versus harmonic loss for given pump power $\chi\beta$ and normalized frequency $\Omega$. Both $S_{X_1+X_2}^{out}$ and $S_{Y_1-Y_2}^{out}$ promptly increase when $\gamma_0$ increases from 0 to 0.02, then they close to SQL. With enough small $\gamma_0$, such as $\gamma_0 = 0.002$, the reflected subharmonic modes from a triply resonating cavity for type II SHG are in an strongly entangled state with the anticorrelation of amplitude quadratures and the correlation of phase quadratures [8] [9].

## 3 Inseparability of amplitude and phase quadratures

The inseparability criterion of EPR entanglement state for continuous variables proposed by Duan [18] is

$$S_{X_1+X_2}^{out}(\Omega) + S_{Y_1-Y_2}^{out}(\Omega) < 2 \quad (15)$$

which is suitable for entangled beams with equal quadrature fluctuations. For the unequal case, a symmetrization procedure has to be carried out, and the criterion becomes of [21]

$$S^{out}_{X_1+X_2}(\Omega)S^{out}_{Y_1-Y_2}(\Omega) < 1. \tag{16}$$

Based on Eqs. (14) we numerically analyzed the dependences of $S^{out}_{X_1+X_2}(\Omega)S^{out}_{Y_1-Y_2}(\Omega)$ on the pump parameters $\sigma$, normalized frequency $\Omega$ and harmonic loss $\gamma_0$. Fig.4 shows the dependences of the entanglement measure on $\gamma_0$ for $\Omega = 0.6$, $\chi\beta = 0.001$, $\gamma = 0.02$ and $\gamma_b = 0.015$. The correlation spectra $S^{out}_{X_1+X_2}(\Omega)S^{out}_{Y_1-Y_2}(\Omega)$ vs. $\sigma$ and $\Omega$ are plotted in figures 5 and 6 respectively, in which we choose $\gamma = 0.02$, $\gamma_b = 0.015$, $\gamma_0 = 0.002$, $\Omega = 0.6$ (Fig.5) and $\sigma = 0.8$ (Fig.6). It is obvious that the inseparability criterion is satisfied in a wide frequency range with reasonable loss parameters. As same as general OPO, the perfect continuous variable entangled state can be obtained in ideal limit.

In summary, we analyzed the continuous-variables entanglement characteristic of the reflected subharmonic field in type II SHG using semiclassical approaches. Dependence of correlation variances on the parameters are caculated. Relative to parametric down-conversion usually used for the entangled-state generation, this scheme is relatively simpler. The reflected pump field from the SHG cavity is an entangled state with the anti-correlated amplitude quadratures and the correlated phase quadratures which is very useful in the quantum information[6][8][9].

Science Foundation supported this work.Science Foundation supported this work.

Fig.1　　Sketch of experimental setup

Fig.2　　The quantum noise spectra of $S^{out}_{X_1+X_2}$ and $S^{out}_{Y_1-Y_2}$ vs. normalized frequency $\Omega$, $\gamma=0.02$, $\gamma_b=0.015$, $\gamma_0=0.002$ and $\sigma=0.8$.

Fig.3　　The quantum noise spectra of $S^{out}_{X_1+X_2}$ and $S^{out}_{Y_1-Y_2}$ vs. harmonic loss $\gamma_0$, $\gamma=0.02$, $\gamma_b=0.015$, $\Omega=0.6$ and $\chi\beta=0.001$.

Fig.4　　The quantum noise spectrum of $S^{out}_{X_1+X_2}(\Omega)S^{out}_{Y_1-Y_2}(\Omega)$ vs. harmonic loss $\gamma_0$, $\Omega=0.6$, $\gamma=0.02$, $\gamma_b=0.015$ and $\chi\beta=0.001$

Fig.5　　The quantum noise spectrum of $S^{out}_{X_1+X_2}(\Omega)S^{out}_{Y_1-Y_2}(\Omega)$ vs. pump parameter $\sigma$, $\gamma=0.02$, $\gamma_b=0.015$, $\gamma_0=0.002$ and $\Omega=0.6$.

Fig.6　　The quantum noise spectrum of $S^{out}_{X_1+X_2}(\Omega)S^{out}_{Y_1-Y_2}(\Omega)$ vs. normalized frequency $\Omega$, $\gamma=0.02$, $\gamma_b=0.015$, $\gamma_0=0.002$ and $\sigma=0.8$.

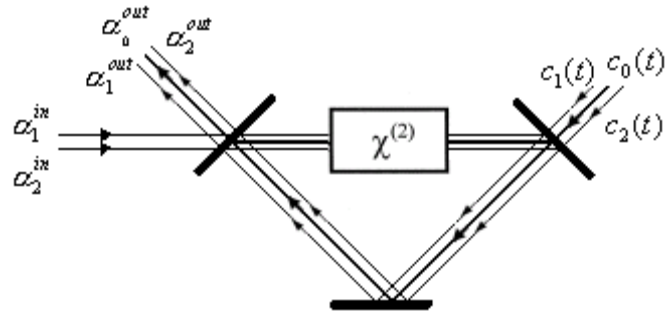

Fig.1 Sketch of experimental setup

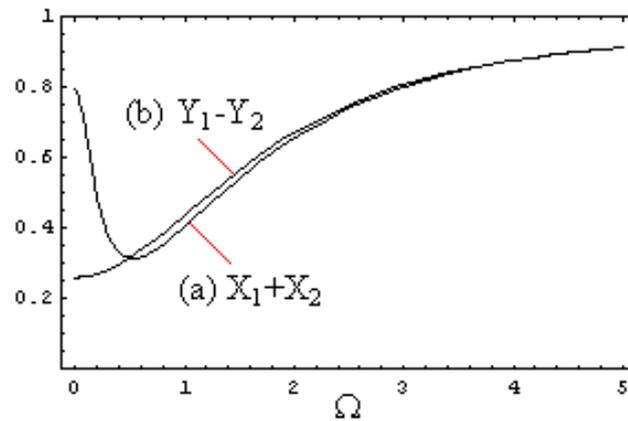

Fig.2 $S^{out}_{X_1+X_2}(\Omega)$ and $S^{out}_{Y_1-Y_2}(\Omega)$ vs. $\Omega$

$\sigma = 0.8$, $\gamma = 0.02$, $\gamma_b = 0.015$, $\gamma_0 = 0.002$

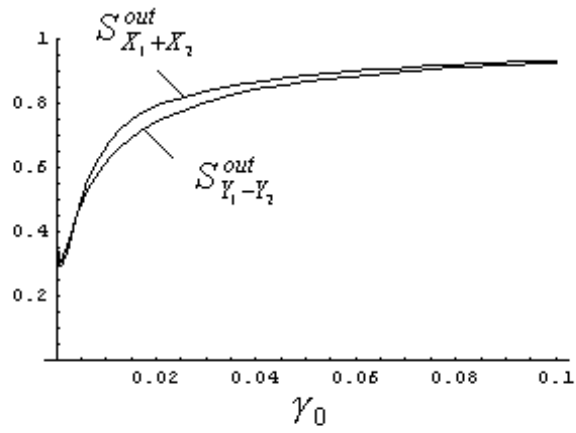

Fig.3 $S^{out}_{X_1+X_2}$ and $S^{out}_{Y_1-Y_2}$ vs. $\gamma_0$

$\Omega = 0.6$, $\gamma = 0.02$, $\gamma_b = 0.015$ and $\chi\beta = 0.001$

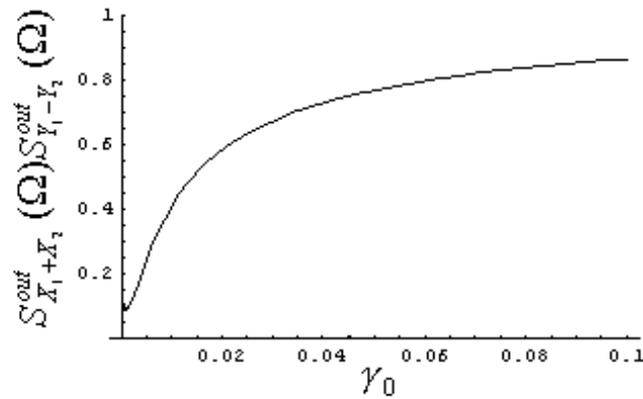

Fig. 4 $S^{out}_{X_1+X_2}(\Omega)S^{out}_{Y_1-Y_2}(\Omega)$ vs. $\gamma_0$

$\Omega = 0.6$, $\gamma = 0.02$, $\gamma_b = 0.015$ and $\chi\beta = 0.001$

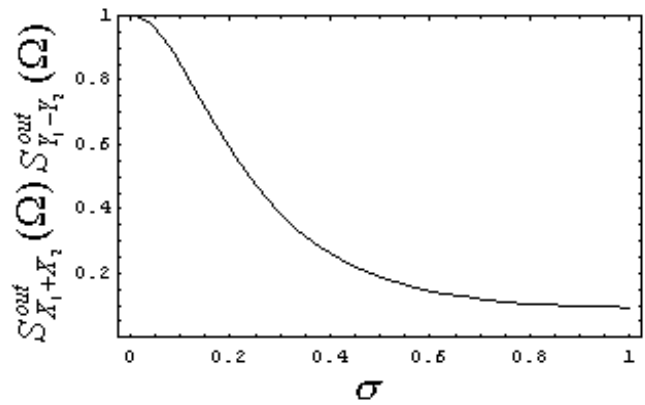

Fig. 5 $S^{out}_{X_1+X_2}(\Omega) S^{out}_{Y_1-Y_2}(\Omega)$ vs. $\sigma$

$\Omega = 0.6$, $\gamma = 0.02$, $\gamma_b = 0.015$, $\gamma_0 = 0.002$

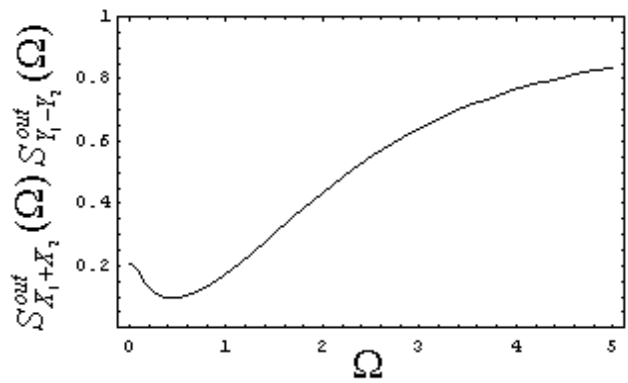

Fig. 6 $S^{out}_{X_1+X_2}(\Omega) S^{out}_{Y_1-Y_2}(\Omega)$ vs. $\Omega$

$\sigma = 0.8$, $\gamma = 0.02$, $\gamma_b = 0.015$, $\gamma_0 = 0.002$